\documentclass[%
 reprint,
superscriptaddress,
 amsmath,amssymb,
 aps,
]{revtex4-2}

\usepackage{microtype}
\usepackage{needspace}
\usepackage{placeins}
\usepackage{physics}
\usepackage{graphicx}
\usepackage{dcolumn}
\usepackage{bm}
\usepackage{hyperref}
\hypersetup{hidelinks}
\usepackage{xcolor}
\usepackage[mathlines]{lineno}

\begin{document}

\preprint{APS/123-QED}

\title{Time-Domain Multiplexing for Resource-Efficient Heralded GKP State Preparation}

\author{Ke-Ming Hu}
\author{Jun-Jie He}
\author{Yu-Ze Zhu}
\author{Pei-Lin Du}
\author{Lin Yang}
\affiliation{TuringQ Co., Ltd., Shanghai 200240, China}

\author{Hao Tang}
\email{htang2015@sjtu.edu.cn}
\affiliation{Center for Integrated Quantum Information Technologies (IQIT), School of Physics and Astronomy and State Key Laboratory of Photonics and Communications, Shanghai Jiao Tong University, Shanghai 200240, China}

\author{Xian-Min Jin}
\email{xianmin.jin@sjtu.edu.cn}
\affiliation{TuringQ Co., Ltd., Shanghai 200240, China}
\affiliation{Center for Integrated Quantum Information Technologies (IQIT), School of Physics and Astronomy and State Key Laboratory of Photonics and Communications, Shanghai Jiao Tong University, Shanghai 200240, China}
\affiliation{Chip Hub for Integrated Photonics Xplore (CHIPX), Shanghai Jiao Tong University, Wuxi 214000, China}
\affiliation{Atomology Co., Ltd., Shanghai 200240, China}

\date{\today}

\begin{abstract}
{Gottesman–Kitaev–Preskill (GKP) states are key resources for fault-tolerant photonic quantum computing, but their conditional preparation via Gaussian boson sampling involves a trade-off between source count and photon-number resolution. Distributing heralding photons across more measured modes reduces the resolution required per mode, yet increases the number of simultaneously operated squeezed sources in fully squeezed spatial architectures. Here, we show that time-domain multiplexing (TDM) can reduce these physical source requirements while preserving GKP state quality. Using an end-to-end differentiable Bargmann framework, we find that TDM circuits reusing one or two squeezed sources achieve fidelities comparable to fully squeezed spatial circuits across several configurations. Selected nine-mode configurations retain fidelities above 0.95 with one photon heralded in each measured mode, mitigating detector resolution demands without a proportional increase in source count. Loss-aware reoptimization further shows that, under the investigated spatial loss models, two-mode TDM maintains fidelities near 0.80 at 1.0~dB per channel, whereas the corresponding spatial benchmarks fall below 0.60. These results demonstrate that temporal reuse provides an effective architectural route for balancing source overhead, detection demands, and state preparation performance in the presence of optical loss.}

\end{abstract}

\maketitle


\section{Introduction}
\label{sec:introduction}

{Continuous-variable (CV) photonic quantum computing provides deterministic Gaussian operations on optical modes~\cite{Weedbrook2012Gaussian}, while fault-tolerant computation requires quantum error correction to protect against noise~\cite{Menicucci2014Fault}. The Gottesman--Kitaev--Preskill (GKP) code encodes logical qubits in periodic {phase space} grids, allowing small displacement errors in both position and momentum quadratures to be corrected~\cite{Gottesman2001Encoding,Grimsmo2021GKP}. Non-Gaussian GKP resource states, combined with Gaussian operations, can support universal and fault-tolerant quantum computation when the squeezing and noise levels meet the requirements of the chosen protocol~\cite{Menicucci2014Fault,Baragiola2019AllGaussian}. Preparing GKP states with sufficient squeezing and fidelity is therefore a key requirement for implementing these photonic computing schemes.}

{Ideal GKP codewords require infinite energy, so physical implementations use finite-energy approximations~\cite{Gottesman2001Encoding,Tzitrin2020Practical,Matsuura2020Equivalence}. Prominent approaches to photonic GKP state preparation include cat breeding and conditioning multimode Gaussian states on photon-number measurement outcomes~\cite{Vasconcelos2010AllOptical,Weigand2018Grid,Su2019Conversion,Tzitrin2020Practical}. Cat-breeding protocols combine interference of Schr\"odinger cat states with homodyne conditioning; suitable measurement-dependent corrections can remove the need for post-selection~\cite{Vasconcelos2010AllOptical,Weigand2018Grid}, although optical loss can limit the quality of the resulting states~\cite{Solodovnikova2025Loss}. In Gaussian boson sampling (GBS) architectures~\cite{Hamilton2017GBS}, squeezed inputs undergo multimode Gaussian interference, and photon-number measurements on selected outputs herald an approximate GKP state in a retained mode~\cite{Su2019Conversion,Tzitrin2020Practical,Larsen2025Integrated}. Related optical proposals combine photon catalysis or photon-number measurements with breeding to prepare approximate GKP states~\cite{Eaton2019Catalysis,Takase2023Synthesizer,Winnel2024Deterministic}. Experimental GKP state preparation has been demonstrated in propagating light and on an integrated photonic platform~\cite{Konno2024Logical,Larsen2025Integrated}. A central architectural challenge is to reduce the number of physical squeezed sources and the photon-number resolution required per measured mode while preserving the quality of the heralded GKP state.}
{GBS-based {GKP state} preparation involves a trade-off between state quality, photon-number resolution, and the number of simultaneously operated squeezed sources. Preparing highly squeezed approximate GKP states can require conditioning on {multiphoton} detection events~\cite{Tzitrin2020Practical,Larsen2025Integrated}. Concentrating these photons in a few measured modes demands high photon-number resolution from individual detectors, and finite detection efficiency affects conditional state preparation~\cite{Quesada2019Realistic,Larsen2025Integrated}. Distributing a fixed total photon number across more measured modes can reduce the largest photon number required in an individual detection outcome. In a spatial architecture with every input squeezed, however, each additional mode requires another simultaneous squeezed source, increasing the demands on synchronization, phase stability, and mode matching~\cite{Larsen2025Integrated}. Time-domain multiplexing (TDM) with optical delay loops reuses sources and optical components across successive time bins~\cite{Motes2014Scalable}. Temporal multiplexing has enabled large continuous-variable cluster states~\cite{Yokoyama2013Cluster}, including two-dimensional resources~\cite{Asavanant2019Cluster,Larsen2019Cluster}, as well as programmable GBS with photon-number-resolving detection~\cite{Madsen2022Advantage}. It therefore allows mode expansion to reduce per-mode photon-number-resolution requirements without a proportional increase in physical source count. Related work has explored TDM GBS architectures through algorithm--hardware co-design for financial graph problems~\cite{He2026Finance}. A recent experiment has also demonstrated a chip-scale space-time multiplexed GBS processor integrating electro-optic modulators and delay lines on thin-film lithium niobate~\cite{Fu2026ChipScale}.}

{Comparing these architectures requires optimizing their circuit parameters against a common objective. Variational optimization has also been used to design heralded non-Gaussian resource-state circuits~\cite{Sabapathy2019Production}. Explicitly constructing the full multimode state in the Fock basis becomes computationally demanding as the number of modes increases~\cite{Su2019Conversion,Motamedi2026Stellar}. We use the Bargmann representation to evaluate the conditional output state without constructing the full multimode Fock tensor~\cite{Miatto2020Fast,Motamedi2026Stellar}. Its integration into an end-to-end differentiable framework enables gradient-based optimization of fidelity, effective squeezing, and heralding probability through a joint objective. This provides a common numerical basis for evaluating spatial and TDM architectures.}

{In this work, we compare spatial and TDM architectures for preparing finite-energy square-lattice GKP states, with heralding measurements detecting a total of $N=8$ or $N=12$ photons. We benchmark TDM circuits using one or two physical squeezed sources against two classes of spatial circuits: those with all input modes squeezed and those with only two squeezed inputs. For several matched photon-number, mode-count, and heralding-pattern configurations, TDM circuits achieve comparable fidelities to fully squeezed spatial circuits, with comparable or higher heralding probabilities in some cases. At $N=8$, selected nine-mode TDM configurations attain fidelities above $0.95$ with one photon detected in each measured mode. With circuit parameters reoptimized at each loss level, the investigated two-mode TDM configurations retain fidelities of approximately $0.80$ at $1.0\,\mathrm{dB}$ per specified {spatial loss} channel, compared with values below $0.60$ for the corresponding spatial benchmarks. This comparison uses the loss placements specified for each architecture. Together, these results show that appropriately configured TDM circuits can combine fewer simultaneous squeezed sources with lower per-mode photon-number-resolution requirements, while achieving higher optimized fidelities under the investigated {spatial loss} models.}


\section{METHOD}
\subsection{End-to-End Optimization Framework}
\label{sec:optimization-framework}
\begin{figure*}[!t]
    \centering
    \includegraphics[width=\linewidth,height=0.70\textheight,keepaspectratio]{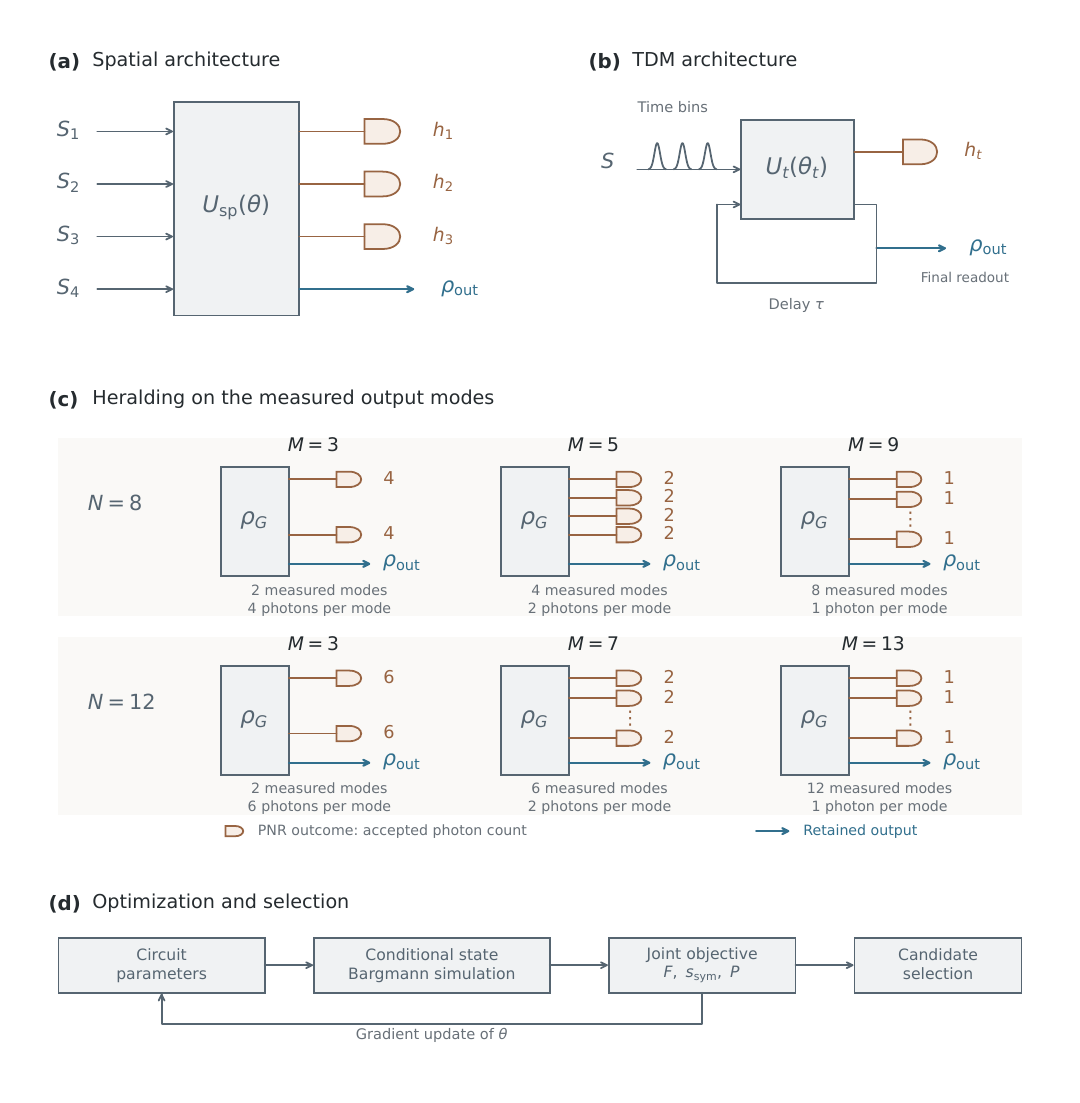}
    \caption{Architecture comparison and conditional {GKP state} optimization. (a) Spatial architecture: parallel squeezed inputs $S_j$ are mixed by a parameterized interferometer $U_{\mathrm{sp}}(\theta)$; photon-number-resolving (PNR) outcomes $h_j$ herald the retained state $\rho_{\mathrm{out}}$. (b) TDM architecture: a squeezed source supplies successive time bins that interact through $U_t(\theta_t)$ and a delay loop of duration $\tau$, with the retained mode extracted at the final readout. Both circuits are schematic. (c) Representative prescribed heralding patterns for total detected photon numbers $N=8$ and $N=12$. Here $\rho_G$ is the Gaussian state before conditioning, and $M$ includes the $M-1$ measured modes and one retained mode. (d) Circuit parameters are optimized by differentiating a joint objective combining fidelity $F$, stabilizer-based effective squeezing $s_{\mathrm{sym}}$, and heralding probability $P$, together with stabilizer penalties, followed by candidate selection. Fidelity is evaluated against the finite-energy GKP $\ket{0}$ target with $\Delta=\kappa=0.36$.}
    \label{fig:protocol}
\end{figure*}

{We use an end-to-end differentiable framework based on DeepQuantum~\cite{he2025deepquantum} to optimize conditional {GKP state} preparation (Fig.~\ref{fig:protocol}). Parameterized photonic circuits and Bargmann {conditional state} calculations are integrated into a common computational graph, allowing gradients of the physical objective to propagate to the circuit parameters. The trainable parameters $\theta$ comprise the squeezing, phase, and interferometric parameters permitted by each architecture.}

{The total mode count $M$ refers to the spatial circuit or the temporally expanded TDM circuit. Gaussian evolution of squeezed-vacuum and vacuum inputs produces the state $\rho_G(\theta)$, which can be pure or mixed in the presence of optical loss. The retained mode is denoted by $o$, and $H$ is the set of $M-1$ measured modes. For ideal PNR detection with prescribed outcome $\boldsymbol h=(h_j)_{j\in H}$ and total detected photon number $N=\sum_{j\in H}h_j$, the unnormalized conditional state, heralding probability, and normalized output state are
\begin{align}
\widetilde\rho_o(\theta\mid\boldsymbol h)
 &= {}_H\!\langle\boldsymbol h|\rho_G(\theta)|\boldsymbol h\rangle_H,
 \label{eq:method-conditional-state}\\
P_{\boldsymbol h}(\theta)
 &=\operatorname{Tr}_o\widetilde\rho_o(\theta\mid\boldsymbol h),\\
\rho_o(\theta\mid\boldsymbol h)
 &=\frac{\widetilde\rho_o(\theta\mid\boldsymbol h)}
 {P_{\boldsymbol h}(\theta)},\qquad P_{\boldsymbol h}>0.
\end{align}
The Bargmann kernel formulation yields the required conditional Fock matrix elements from the Gaussian covariance matrix and displacement vector without constructing the full $M$-mode Fock density tensor (Appendix~\ref{app:bargmann}). The output state's quadrature structure is characterized by the marginal probability densities $p_q(q)=\langle q|\rho_o|q\rangle$ and $p_p(p)=\langle p|\rho_o|p\rangle$.}

{We use a normalized finite-energy approximation to the square-lattice GKP $\ket{0}$ state as the common target~\cite{Gottesman2001Encoding,Tzitrin2020Practical}. We use dimensionless quadratures, with $\langle q|q'\rangle=\delta(q-q')$ and $\langle p|q\rangle=e^{-ipq}/\sqrt{2\pi}$. These correspond to the numerical quadratures of Appendix~\ref{app:bargmann} divided by $\sqrt{\hbar}$. For $\Delta,\kappa>0$, our target is defined by
\begin{align}
|G_{\Delta,\kappa}\rangle
 &=\int_{\mathbb R}\psi_{\Delta,\kappa}(q)|q\rangle\,dq,
 \label{eq:gkp-target-state}\\
\psi_{\Delta,\kappa}(q)
 &=\frac{g_{\Delta,\kappa}(q)}{\sqrt{\mathcal Z_{\Delta,\kappa}}},\\
g_{\Delta,\kappa}(q)
 &=\sum_{s\in\mathbb Z}e^{-2\pi\kappa^2s^2}
   \exp\!\left[-\frac{(q-2s\sqrt{\pi})^2}{2\Delta^2}\right],
 \label{eq:gkp-target-comb}\\
\mathcal Z_{\Delta,\kappa}
 &=\int_{\mathbb R}|g_{\Delta,\kappa}(q)|^2\,dq.
 \label{eq:gkp-target-normalization}
\end{align}
The momentum wavefunction is fixed by the same state through the Fourier transform,
\begin{equation}
\begin{aligned}
\widetilde\psi_{\Delta,\kappa}(p)
 &=\frac{1}{\sqrt{2\pi}}\int_{\mathbb R}
       e^{-ipq}\psi_{\Delta,\kappa}(q)\,dq\\
 &=\frac{\Delta}{\sqrt{\mathcal Z_{\Delta,\kappa}}}
       e^{-\Delta^2p^2/2}
       \sum_{s\in\mathbb Z}e^{-2\pi\kappa^2s^2}
       e^{-i2s\sqrt{\pi}p}.
\end{aligned}
\label{eq:gkp-target-momentum}
\end{equation}
In this definition, $\Delta$ sets the width of each Gaussian amplitude peak, while $\kappa$ controls the weights at the lattice sites. Throughout the comparisons, we fix $\Delta=\kappa=0.36$ and write $|G\rangle\equiv|G_{0.36,0.36}\rangle$.}

{For both ideal-circuit and loss-aware optimization, circuit parameters minimize the joint loss
\begin{equation}
\begin{aligned}
\mathcal{L}(\theta)={}&-\alpha_1F-\alpha_2s_{\mathrm{sym}}\\
&-\alpha_3\log_{10}\!\left[\max(P,10^{-16})\right]\\
&+\alpha_4C_{\mathrm{phase}}+\alpha_5C_{\mathrm{balance}},
\end{aligned}
\label{eq:plotted-optimization-loss}
\end{equation}
where $P=P_{\boldsymbol h}(\theta)$ and $F=\langle G|\rho_o|G\rangle$ is the squared fidelity to the target defined above. The same weights are used for all compared architectures and loss levels,
\begin{equation}
(\alpha_1,\alpha_2,\alpha_3,\alpha_4,\alpha_5)
=(5,\,0.2,\,0.1,\,2,\,0.5).
\label{eq:optimization-weights}
\end{equation}

For the same quadrature convention, the stabilizers are $S_q=e^{i2\sqrt{\pi}q}$, $S_p=e^{-i2\sqrt{\pi}p}$, with expectations $z_\mu=\operatorname{Tr}(\rho_o S_\mu)$ for $\mu\in\{q,p\}$. The magnitudes $|z_q|$ and $|z_p|$ measure the concentration of the respective quadrature distributions after reduction modulo $\sqrt{\pi}$~\cite{Weigand2019Modular}. We use the vacuum-normalized effective variances and symmetric effective squeezing
\begin{equation}
v_\mu=-\frac{\ln|z_\mu|}{\pi},\qquad
s_{\mathrm{sym}}=-10\log_{10}\!\left(\frac{v_q+v_p}{2}\right).
\label{eq:method-effective-squeezing}
\end{equation}
Vacuum gives $v_q=v_p=1$, so $s_{\mathrm{sym}}$ measures effective squeezing relative to vacuum in dB. The stabilizer penalties are
\begin{align}
C_{\mathrm{phase}}&=\frac{|z_q-1|^2+|z_p-1|^2}{2},\\
C_{\mathrm{balance}}&=|z_q-z_p|.
\label{eq:method-stabilizer-penalties}
\end{align}
The first penalty favors stabilizer expectations close to unity in both phase and magnitude; the second penalizes differences between the two complex expectations.}

\begin{figure*}[!t]
    \centering
    \includegraphics[width=\textwidth,height=0.70\textheight,keepaspectratio]{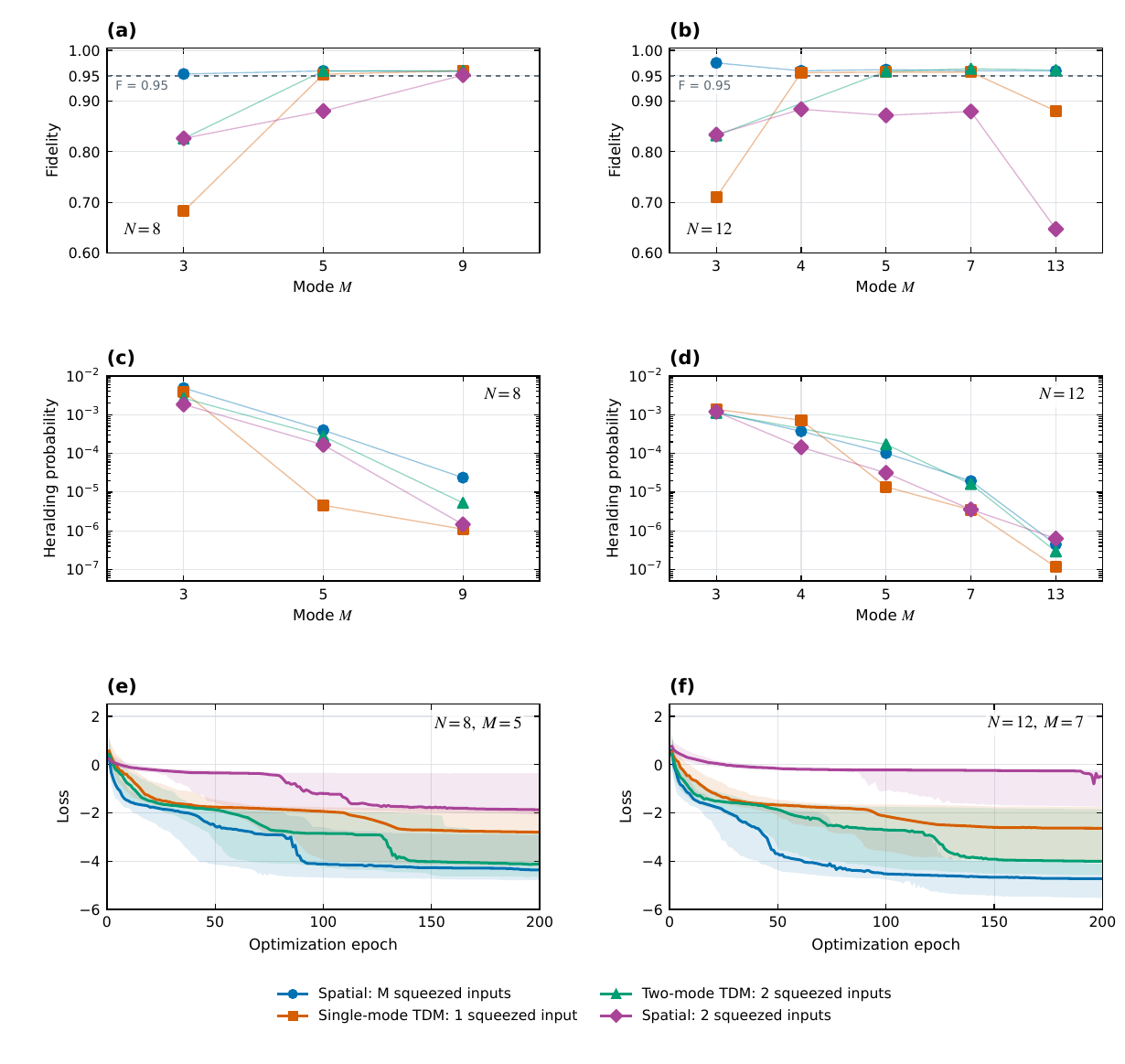}
    \caption{Optimization convergence and final results by expanded mode count. (a,b) Final fidelity and (c,d) heralding probability for $N=8$ and $N=12$, respectively. (e,f) Matched-configuration loss curves for $(N,M)=(8,5)$ and $(12,7)$. Loss is $\mathcal{L}=-\mathrm{score}$ from the archived joint-objective optimizer; lower is better. Solid convergence curves are medians over seeds 0--99, with 25th--75th percentile bands, using the same objective weights and 200 epochs at training cutoff 30. Final $F$ and $P$ are evaluated at cutoff 80. Within each archived configuration and batch, select the largest $P$ with $F>0.95$; otherwise select the largest $P$ among the five highest-fidelity candidates. Repeated configurations retain the highest-$P$ passing batch representative. Final selections use the archived search budgets of 100--300 seeds per batch, rather than the common 100-seed subset used for convergence statistics. Colors and marker shapes identify architectures. The TDM legend entries count squeezed inputs per time step, supplied by one or two reused physical sources; they do not count all squeezed temporal inputs in the expanded circuit. Figure~\ref{fig:appendix-circuits-loss} shows the representative spatial and TDM circuit structures and the one-step temporal expansions. All selected candidates use filled markers, including those below the fidelity threshold. Markers denote available configurations; connecting lines guide the eye.}
    \label{fig:optimization-results}
\end{figure*}

{Optimization is performed from multiple random initializations using the joint loss above. Candidate selection depends on the comparison: Fig.~\ref{fig:optimization-results} prioritizes heralding probability subject to $F>0.95$, with the fallback and batch-selection rules specified in its caption. For Fig.~\ref{fig:mode-matched-fidelity-loss}, parameters are independently reoptimized at each loss level, and the highest verified fidelity is selected among the saved maximum-fidelity checkpoints from 100 runs. Thus, the selection criterion is distinct from the objective used for gradient updates.}

{Circuit parameters are optimized with a {retained mode} Fock cutoff of $C=30$. Final ideal-circuit results are evaluated at $C=80$, while loss-aware results are checked at larger cutoffs to ensure convergence. Numerical details are provided in Appendix~\ref{app:bargmann}.}

\subsection{Spatial and {Time-Domain Multiplexed} Architectures}
{We compare spatial circuits with either $M$ squeezed-vacuum inputs or two squeezed-vacuum inputs and vacuum in the remaining $M-2$ modes. Both configurations use a Clements interferometer~\cite{Clements2016Optimal}, consisting of a rectangular mesh of Mach--Zehnder interferometers (MZIs) followed by output phase shifts. The trainable parameters are the squeezing amplitudes and phases, MZI angles, and output phases. PNR measurements on $M-1$ output modes herald the retained state. Figure~\ref{fig:appendix-circuits-loss}(a) illustrates a four-mode layout; the loss channels shown there are described in Appendix~\ref{app:spatial-loss-circuits}.}

{The TDM architectures reuse one or two physical squeezed sources to supply pulses at successive time steps. In the single-mode circuit, each incoming pulse couples through an MZI to a stored mode in a loop with a single {time bin} delay [Fig.~\ref{fig:appendix-circuits-loss}(b)]. The two-mode circuit additionally mixes the two pulse streams through MZIs before and after the coupling to the stored mode [Fig.~\ref{fig:appendix-circuits-loss}(c)]. Including the initially vacant storage mode, temporal expansion over $T$ steps gives $M=T+1$ modes for single-mode TDM and $M=2T+1$ for two-mode TDM. The squeezing amplitudes and phases and MZI angles can vary at each time step, while optimization preserves the prescribed TDM connectivity. Source reuse therefore requires programmable settings across time bins. The final loop mode is retained, and all other expanded modes are measured with the prescribed PNR outcomes.}

Optical attenuation is modeled via a beam-splitter loss channel coupled to a vacuum environment (Appendix~\ref{app:optical-loss}). Depending on its circuit placement, this model describes either spatial loss in the output waveguides prior to detection or delay line loss incurred during each loop traversal. The architecture-dependent channel configurations are detailed in Appendix~\ref{app:spatial-loss-circuits}.

{For each total detected photon number $N=8$ or $N=12$, we compare mode counts $M$ and prescribed heralding patterns $\boldsymbol h$ satisfying $\sum_{j\in H}h_j=N$. Representative patterns are shown in Fig.~\ref{fig:protocol}(c). Distributing the photons over more measured modes can reduce $\max_{j\in H}h_j$, lowering the photon-number resolution required for an individual accepted outcome. Each pattern is fixed during optimization, and only the circuit parameters are varied.}

\section{RESULTS}

{Figures~\ref{fig:optimization-results}(a--d) summarize the performance of the investigated architectures for total detected photon numbers $N=8$ and $N=12$. The fully squeezed spatial architecture achieves $F>0.95$ across all displayed mode counts. For $N=8$ [Fig.~\ref{fig:optimization-results}(a)], the selected single- and two-mode TDM circuits at $M=9$ attain fidelities of $0.9600$ and $0.9599$, respectively, comparable to the nine-source spatial benchmark of $0.9582$ while using only one or two physical squeezed sources. In this configuration, the heralding pattern specifies one photon in each measured mode ($h_j=1$), reducing the photon-number resolution required for the accepted outcomes. For $N=12$ [Fig.~\ref{fig:optimization-results}(b)], the corresponding single-photon heralding configuration at $M=13$ yields $F=0.9611$ for two-mode TDM, close to the spatial benchmark of $0.9595$.}

{These comparable fidelities are accompanied by lower heralding probabilities in the single-photon heralding configurations [Figs.~\ref{fig:optimization-results}(c,d)]. For $N=8$ and $M=9$, single- and two-mode TDM yield probabilities of $1.10\times10^{-6}$ and $5.27\times10^{-6}$, respectively, compared with $2.37\times10^{-5}$ for the spatial benchmark. For $N=12$ and $M=13$, the two-mode TDM probability is $2.94\times10^{-7}$. The restricted connectivity of the investigated single-loop circuits limits the accessible interferometer transformations relative to a universal Clements mesh and may contribute to these probability differences. Within the investigated configurations, these results illustrate a trade-off among physical squeezed source count, per-mode detection requirements, and heralding probability.}

{A complementary comparison fixes the number of physical squeezed sources at two. For $(N,M)=(8,5)$ and $(12,7)$, the selected states prepared by the two-mode TDM architecture achieve fidelities of $0.9595$ and $0.9638$, respectively, compared with $0.8802$ and $0.8793$ for the corresponding spatial circuits with two squeezed inputs. Although the physical source counts are equal, the available multimode squeezing resources differ. The spatial circuit combines two squeezed inputs with $M-2$ vacuum inputs, and its passive Clements interferometer redistributes the squeezing without introducing additional squeezed inputs. Two-mode TDM instead supplies two squeezed pulses at each time step, yielding $2T$ squeezed inputs in the temporally expanded circuit with $M=2T+1$. This resource difference provides a physical basis for interpreting the improved {conditional state} fidelities. Figures~\ref{fig:optimization-results}(e,f) further show that, over 100 runs using common seed labels, two-mode TDM attains a lower final median joint loss than the two-squeezer spatial circuit, while remaining above the fully squeezed spatial benchmark. Within the investigated configurations and training budget, temporal reuse thus enables a small number of physical sources to supply a larger set of squeezed inputs and approach the performance of fully squeezed spatial circuits.}

\begin{figure*}[!t]
    \centering
    \includegraphics[width=\textwidth,height=0.70\textheight,keepaspectratio]{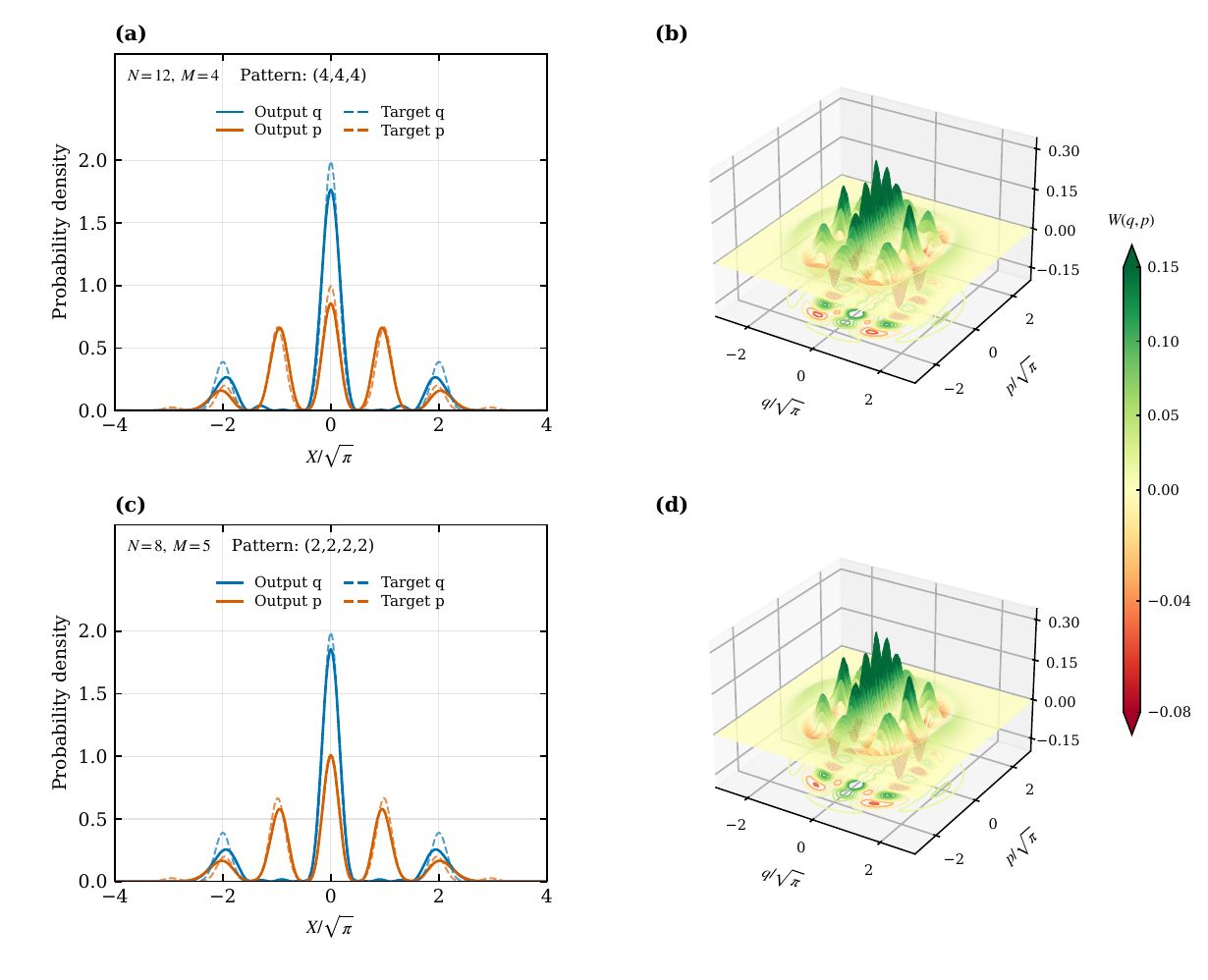}
    \caption{{Selected states prepared by the single- and two-mode TDM architectures. For each architecture, the state with the highest heralding probability among all archived candidates satisfying $F>0.95$ is shown. (a,b) Marginal probability densities and Wigner function for single-mode TDM with $N=12$, $M=4$, and $\boldsymbol h=(4,4,4)$, yielding $F=0.9560$ and $P=7.22\times10^{-4}$. (c,d) The corresponding distributions for two-mode TDM with $N=8$, $M=5$, and $\boldsymbol h=(2,2,2,2)$, yielding $F=0.9595$ and $P=2.78\times10^{-4}$. Solid curves show the output quadrature probability densities, and dashed curves show those of the finite-energy GKP target with $\Delta=\kappa=0.36$. In (a,c), $X$ denotes either $q$ or $p$; the densities include the Jacobian associated with the scaled coordinate $X/\sqrt{\pi}$ and are normalized over the full quadrature domain. Both Wigner plots use $[q,p]=i$, identical axes, and a common color scale; surface heights retain the actual values beyond the color range.}}
    \label{fig:gkp}
\end{figure*}
{Figure~\ref{fig:gkp} characterizes the {phase space} structure of states prepared by the single- and two-mode TDM architectures. For each architecture, the state with the highest heralding probability among archived candidates satisfying $F>0.95$ is selected. The quadrature distributions in panels (a,c) exhibit the characteristic comb structure of the finite-energy GKP target, whose position and momentum peaks are spaced by $2\sqrt{\pi}$ and $\sqrt{\pi}$, respectively. The corresponding Wigner functions in panels (b,d) reveal a lattice of positive peaks interspersed with negative interference regions. These negative regions establish the nonclassical character of the conditional output states and complement the grid structure visible in the quadrature distributions. Together, these features show that conditional photon-number measurements in the investigated TDM circuits produce high-fidelity states with the characteristic {phase space} structure of the target GKP state.}

\begin{figure*}[!t]
    \centering
    \includegraphics[width=\textwidth]{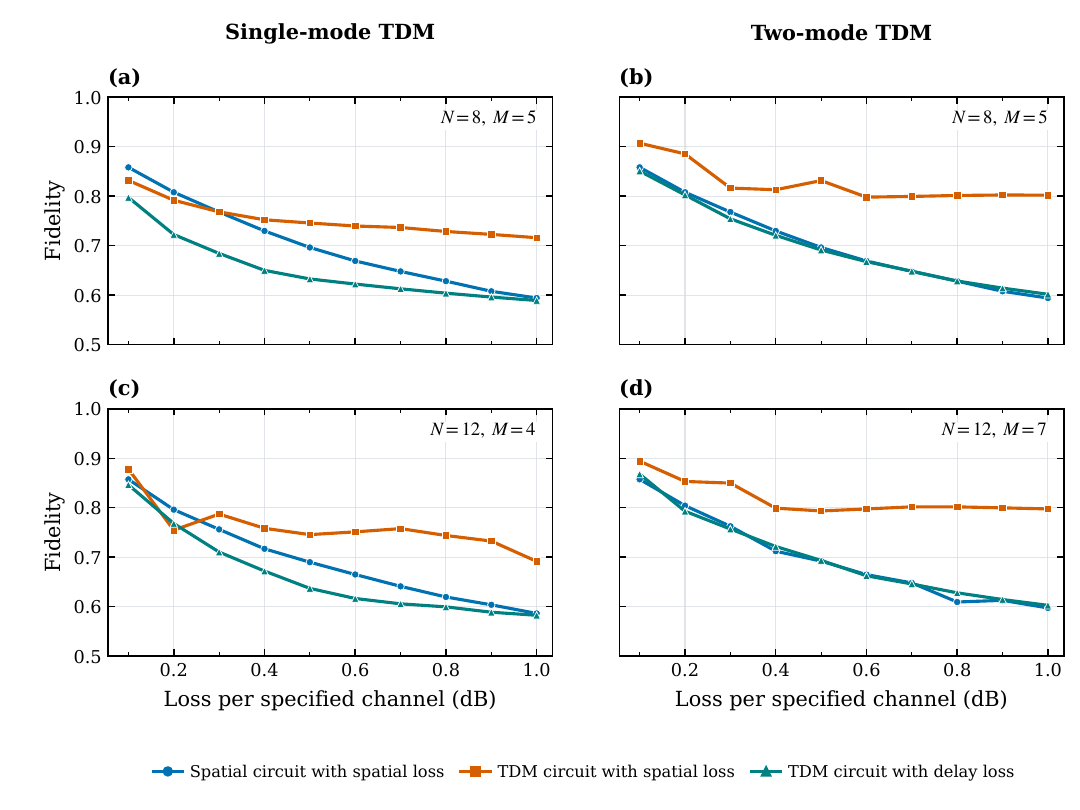}
    \caption{Fidelity after loss-aware optimization of spatial and TDM circuits matched in detected photon number $N$, expanded mode count $M$, and heralding pattern. Panels (a,b) use $(N,M)=(8,5)$ with $\boldsymbol h=(2,2,2,2)$; (c) uses $(12,4)$ with $\boldsymbol h=(4,4,4)$; (d) uses $(12,7)$ with $\boldsymbol h=(2,2,2,2,2,2)$. The spatial circuits use $M$ squeezed inputs; the left and right columns compare them with single- and two-mode TDM, respectively. At each loss value, parameters are optimized using the joint objective in Eq.~\eqref{eq:plotted-optimization-loss}, with common seed labels 0--99 and 200 Adam updates per seed. Each point is the highest verified fidelity among the saved maximum-fidelity checkpoints of these runs. Spatial-circuit loss acts on all terminal modes, including the retained mode; TDM spatial loss acts on outgoing physical rails at each time step, excluding the final retained loop mode; delay loss acts per loop traversal. Only one loss type is active at a time, and equal dB values do not imply equal cumulative attenuation or identical loss locations.}
    \label{fig:mode-matched-fidelity-loss}
\end{figure*}
{Figure~\ref{fig:mode-matched-fidelity-loss} compares the highest verified fidelities found after independently optimizing circuit parameters at each specified loss level using the same joint objective and training budget. Under spatial loss, two-mode TDM maintains higher fidelities than the fully squeezed spatial benchmarks throughout the sampled range in panels (b,d), with the curves approaching a plateau near $F=0.80$ at higher losses. At $1.0\,\mathrm{dB}$, the fidelities are $0.8020$ and $0.7976$, compared with $0.5939$ and $0.5973$ for the spatial circuits, giving absolute fidelity advantages of $0.2080$ and $0.2002$, respectively. Single-mode TDM likewise exceeds the spatial benchmarks at higher {spatial loss} values in panels (a,c). {Under delay loss, two-mode TDM remains close to the {spatial loss} benchmarks in panels (b,d), whereas single-mode TDM yields lower fidelities in panels (a,c), with more pronounced differences at intermediate loss values.} The interpretation depends on the loss placement: the spatial model attenuates all terminal modes, including the retained output, whereas TDM spatial loss acts on outgoing measured rails and excludes the final retained loop mode. The observed advantage therefore reflects circuit connectivity, channel placement, and parameter adaptation together. By varying the number of time steps and local circuit structure, TDM provides flexibility in the effective spatial mode count while reusing a small number of squeezed sources. These results show that appropriately configured TDM circuits can combine reduced physical source requirements with higher optimized fidelities under the specified {optical loss} models.}

\section{CONCLUSION}

{Using an end-to-end differentiable Bargmann framework, we benchmarked spatial and time-domain multiplexed architectures for finite-energy GKP state preparation. Reusing one or two physical squeezed sources yields fidelities comparable to those of fully squeezed spatial circuits in several configurations. In the four-mode example, single-mode TDM reduces simultaneously operated squeezed sources by 75\% while achieving a higher heralding probability. Expanding the temporal mode count distributes multiphoton heralding conditions across single-photon projections in multiple measured modes, reducing the photon-number resolution required for accepted outcomes. However, the investigated single-photon heralding configurations also exhibit lower success probabilities, revealing a trade-off among source count, detection requirements, and preparation performance. Loss-aware reoptimization further shows that, under the specified spatial loss models, two-mode TDM maintains fidelities near $F=0.80$ at higher losses and exceeds the corresponding spatial benchmarks by more than $0.20$ at $1.0\,\mathrm{dB}$. These results demonstrate that TDM can reduce simultaneous physical source requirements while exploiting temporally adjustable parameters to achieve higher output fidelities under the investigated loss conditions, offering a route to resource-efficient state preparation in the presence of optical loss.}

{These results were obtained with elementary single delay loop TDM architectures, motivating further exploration of richer temporal connectivity. Multiloop and recirculating networks~\cite{Motes2014Scalable,Takeda2017Universal}, together with generalized Green Machine architectures~\cite{Basani2026Hardware}, warrant investigation to determine whether additional connectivity can improve preparation performance and reduce pulse-to-pulse modulation requirements by allowing fixed squeezing amplitudes and phases, while accounting for additional loss and control overhead~\cite{Motes2015Loss}. At the system level, the fidelities and heralding probabilities reported here provide quantitative inputs for estimating the multiplexing scale and optical buffering requirements of GKP state factories~\cite{Bourassa2021Blueprint,Tzitrin2021Static,AghaeeRad2025Scaling}. Taken together, these results establish time-domain multiplexed Gaussian boson sampling for GKP state preparation (TGBS-GKP) as a resource-efficient approach that combines physical source reuse with circuit optimization to balance detection requirements and output state quality under optical loss.}

\begin{acknowledgments}
This research is supported by the National Key R\&D Program of China (Grants No.~2024YFB4504005, No.~2024YFB4504000, and No.~2024YFA1409300);
National Natural Science Foundation of China (NSFC) (Grants No.~62235012, No.~12304342, No.~12574549, No.~12574542, No.~125B1033, No.~T2600595, and No.~62605257);
Quantum Science and Technology---National Science and Technology Major Project (Grants No.~2021ZD0301500 and No.~2021ZD0300700);
Science and Technology Commission of Shanghai Municipality (STCSM) (Grants No.~2019SHZDZX01, No.~24ZR1438700, No.~24ZR1430700, and No.~24LZ1401500);
Shanghai Municipal Commission of Economy and Informatization (No.~2025-GZL-RGZN-02047);
and Frontier Technologies R\&D Program of Jiangsu (Grant No.~SBF20250000094).
X.-M.J. acknowledges additional support from a Shanghai talent program and support from Zhiyuan Innovative Research Center of Shanghai Jiao Tong University.
Hao Tang acknowledges an open research fund from State Key Laboratory of Photonics and Communications and additional support from Yangyang Development Fund.
\end{acknowledgments}

\begin{figure*}[!t]
    \centering
    \includegraphics[width=\textwidth]{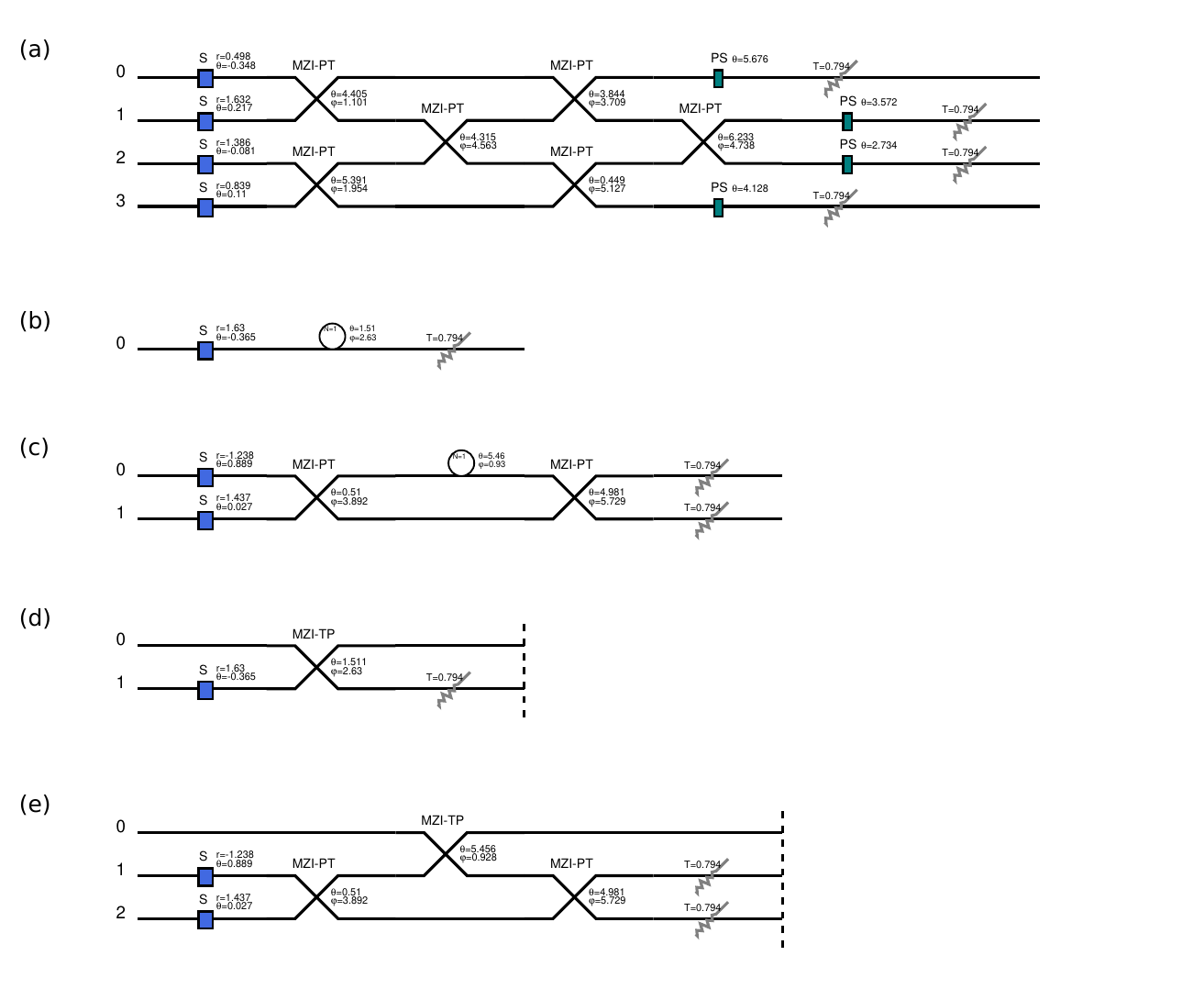}
    \caption{Representative circuit architectures with $1\,\mathrm{dB}$ spatial loss: (a) four-mode spatial GBS with four squeezed inputs; (b,c) local single- and two-mode TDM circuits; (d) spatial expansion of the single-mode TDM unit in (b) over one time step ($T=1$), containing two modes; (e) spatial expansion of the two-mode TDM unit in (c) over one time step ($T=1$), containing three modes. In (d,e), mode 0 is the initially vacant storage mode and is retained, while modes 1 and, in (e), 2 receive squeezed inputs and form the measured outputs. Spatial loss acts on all terminal modes in (a), but only on the outgoing measured modes in the TDM circuits. The TDM names refer to the physical input modes before expansion. Delay line loss is zero; the loop label in (b,c) denotes a single time bin delay.}
    \label{fig:appendix-circuits-loss}
\end{figure*}

\appendix
\Needspace{8\baselineskip}
\section{Circuit architectures and loss modeling}
\label{app:spatial-loss-circuits}

{This appendix presents the circuit layouts and {loss channel} placements for the architectures investigated in this work. Figure~\ref{fig:appendix-circuits-loss} illustrates representative circuit parameters from ideal-circuit optimizations at $N=12$: (a) four-mode spatial GBS with universal Clements mixing and (b) single-mode TDM, both with $M=4$; and (c) two-mode TDM with $M=7$. Panels (b,c) show the local coupling units with parameters from the first time step of the selected candidates. Panel (d) is the spatially unfolded representation of the single-mode TDM unit in panel (b), and panel (e) is that of the two-mode TDM unit in panel (c). Both spatial expansions correspond to one time step ($T=1$) and use the same circuit parameters as their respective local units. They contain $M=2$ and $M=3$ modes, respectively, in accordance with $M=T+1$ and $M=2T+1$. The expansion makes the initially vacant storage mode explicit as mode 0, coupled to one or two incoming squeezed modes. The storage output is retained, while the other outputs are measured. These one-step diagrams illustrate the connectivity and loss placement rather than the full $M=4$ and $M=7$ optimized circuits. Each illustrated {spatial loss} block introduces $1\,\mathrm{dB}$ of attenuation, corresponding to an intensity transmission of $\eta=10^{-1/10}\approx0.7943$ (labeled $T=0.794$ in the diagrams), with {delay line} loss set to zero. These blocks illustrate loss placement; the displayed parameters are not the results of reoptimization at $1\,\mathrm{dB}$.}

{In the spatial architecture [Fig.~\ref{fig:appendix-circuits-loss}(a)], spatial loss is applied uniformly to all terminal modes before detection, including the retained output mode. In the TDM architectures [Figs.~\ref{fig:appendix-circuits-loss}(b--e)], spatial loss is introduced at each time step on the outgoing physical rails before measurement, representing attenuation in the output and detector-coupling paths; the final retained loop mode is excluded from these {spatial loss} channels. {Delay line} loss is introduced independently within the storage loop, with a constant transmission per traversal, so that repeated traversals produce multiplicative attenuation. In Fig.~\ref{fig:mode-matched-fidelity-loss}, the two loss mechanisms are evaluated separately, with circuit parameters reoptimized at each specified loss value using the joint objective and weights in Eqs.~\eqref{eq:plotted-optimization-loss} and \eqref{eq:optimization-weights}. Each plotted point is the highest verified fidelity among the saved maximum-fidelity checkpoints from 100 optimization runs. The beam-splitter model underlying these loss channels is described in Appendix~\ref{app:optical-loss}.}

\section{Heralded-state simulation using the Husimi Q function and Bargmann representation}
\label{app:bargmann}

Fock probabilities and matrix elements of Gaussian states can be evaluated using hafnians and multivariate Hermite polynomials~\cite{Gupt2019Walrus}. We describe a covariance-based Bargmann method for simulating single-mode non-Gaussian states heralded by photon-number-resolving (PNR) measurements on multimode Gaussian states. Gaussian evolution is represented by first moments and covariance matrices; conditional Fock matrix elements are then extracted from a Gaussian density-operator generating function. This formulation accommodates both pure Gaussian states and Gaussian mixed states produced by loss, without introducing an explicit Fock truncation for environmental modes~\cite{Motamedi2026Stellar,Weedbrook2012Gaussian,Su2019Conversion,Quesada2019Realistic}.

\subsection{{Coherent state} projections and Bargmann kernels}

For an $M$-mode system, normalized coherent states satisfy
\begin{equation}
 |\boldsymbol\alpha\rangle=e^{-\|\boldsymbol\alpha\|^2/2}
 \sum_{\boldsymbol n\geq0}\frac{\boldsymbol\alpha^{\boldsymbol n}}{\sqrt{\boldsymbol n!}}|\boldsymbol n\rangle,
 \label{eq:b-coherent}
\end{equation}
where $\boldsymbol n!=\prod_j n_j!$ and $\boldsymbol\alpha^{\boldsymbol n}=\prod_j\alpha_j^{n_j}$. The physical amplitudes obey $\hat a_j|\boldsymbol\alpha\rangle=\alpha_j|\boldsymbol\alpha\rangle$. For $|\psi\rangle=\sum_{\boldsymbol n}\psi_{\boldsymbol n}|\boldsymbol n\rangle$, define
\begin{equation}
 F_\psi(\boldsymbol z)=e^{\|\boldsymbol z\|^2/2}\langle\boldsymbol z^*|\psi\rangle
 =\sum_{\boldsymbol n\geq0}\psi_{\boldsymbol n}\frac{\boldsymbol z^{\boldsymbol n}}{\sqrt{\boldsymbol n!}}.
 \label{eq:b-pure}
\end{equation}
Thus $\boldsymbol z=\boldsymbol\alpha^*$ on a physical {coherent state} projection, while $\boldsymbol z$ is subsequently treated as an independent analytic coordinate. The {coherent state} normalization factor cancels, leaving a holomorphic function whose normalized Taylor coefficients are Fock amplitudes. Creation and annihilation act as
\begin{equation}
 F_{\hat a_j^\dagger\psi}=z_jF_\psi,\qquad
 F_{\hat a_j\psi}=\partial_{z_j}F_\psi.
 \label{eq:b-operators}
\end{equation}

For a density operator, introduce two independent sets of complex variables:
\begin{equation}
 \begin{aligned}
 \mathcal K_\rho(\boldsymbol z,\boldsymbol w)
 &=e^{(\|\boldsymbol z\|^2+\|\boldsymbol w\|^2)/2}
 \langle\boldsymbol z^*|\rho|\boldsymbol w\rangle\\
 &=\sum_{\boldsymbol n,\boldsymbol m\geq0}\rho_{\boldsymbol n,\boldsymbol m}
 \frac{\boldsymbol z^{\boldsymbol n}\boldsymbol w^{\boldsymbol m}}
 {\sqrt{\boldsymbol n!\boldsymbol m!}}.
 \end{aligned}
 \label{eq:b-density-kernel}
\end{equation}
We order the variables as $\boldsymbol\xi=(\boldsymbol z,\boldsymbol w)^{\mathsf T}$, with $\boldsymbol z$ and $\boldsymbol w$ corresponding to row and column indices, respectively. The physical {coherent state} diagonal is obtained at $(\boldsymbol z,\boldsymbol w)=(\boldsymbol\alpha^*,\boldsymbol\alpha)$, but the variables remain independent when extracting coefficients:
\begin{equation}
 \rho_{\boldsymbol n,\boldsymbol m}
 =\left.\frac{\partial_{\boldsymbol z}^{\boldsymbol n}\partial_{\boldsymbol w}^{\boldsymbol m}\mathcal K_\rho}
 {\sqrt{\boldsymbol n!\boldsymbol m!}}\right|_{\boldsymbol z=\boldsymbol w=0}.
 \label{eq:b-elements}
\end{equation}
For a pure state, $\mathcal K_\rho(\boldsymbol z,\boldsymbol w)=F_\psi(\boldsymbol z)F_\psi^\#(\boldsymbol w)$, where $\#$ conjugates coefficients but not formal variables. A mixed-state kernel generally does not factorize in this way. The $2M$ variables describe two {density matrix} indices per mode, not additional physical modes.

\subsection{Gaussian density kernels from the Husimi Q function}

We use $\hat{\boldsymbol R}=(\hat x_1,\ldots,\hat x_M,\hat p_1,\ldots,\hat p_M)^{\mathsf T}$, with $[\hat R_j,\hat R_k]=i\hbar\Omega_{jk}$ and $\Omega=\left(\begin{smallmatrix}0&I\\-I&0\end{smallmatrix}\right)$. Define
\begin{equation}
 \begin{gathered}
 d_j=\langle\hat R_j\rangle,\qquad
 \Sigma_{jk}=\tfrac12\langle\{\hat R_j-d_j,\hat R_k-d_k\}\rangle,\\
 \hat a_j=(\hat x_j+i\hat p_j)/\sqrt{2\hbar}.
 \end{gathered}
 \label{eq:b-moments}
\end{equation}
The vacuum covariance is $\hbar I/2$. We retain $\hbar$ in the derivation and use $\hbar=2$ numerically. Loss is applied to the moments before constructing the kernel. For example, independent {pure loss} channels with vacuum environments give
\begin{equation}
 d'=Ld,\qquad \Sigma'=L\Sigma L^{\mathsf T}+\frac{\hbar}{2}(I-LL^{\mathsf T}),
 \label{eq:b-loss}
\end{equation}
where $L=\operatorname{diag}(\sqrt{\eta_1},\ldots,\sqrt{\eta_M},\sqrt{\eta_1},\ldots,\sqrt{\eta_M})$ and $\eta_j$ is the intensity transmission. Each channel is applied at its specified circuit location.

The Husimi function $Q_\rho(\boldsymbol\alpha)=\pi^{-M}\langle\boldsymbol\alpha|\rho|\boldsymbol\alpha\rangle$ satisfies
\begin{equation}
 \mathcal K_\rho(\boldsymbol\alpha^*,\boldsymbol\alpha)
 =\pi^M e^{\|\boldsymbol\alpha\|^2}Q_\rho(\boldsymbol\alpha).
 \label{eq:b-q-kernel}
\end{equation}
Its expansion in $(\boldsymbol\alpha^*)^{\boldsymbol n}\boldsymbol\alpha^{\boldsymbol m}$ determines both diagonal and off-diagonal Fock matrix elements. For a Gaussian state, let
\begin{equation}
 \begin{gathered}
 \Gamma=\Sigma/\hbar+I/2,\qquad \delta=d/\sqrt\hbar,\\
 r_\alpha=\sqrt{2\hbar}\begin{pmatrix}\operatorname{Re}\boldsymbol\alpha\\\operatorname{Im}\boldsymbol\alpha\end{pmatrix}.
 \end{gathered}
 \label{eq:b-gamma}
\end{equation}
The Gaussian overlap with a coherent state yields
\begin{equation}
 Q_\rho(\boldsymbol\alpha)=
 \frac{\exp[-\tfrac12(r_\alpha/\sqrt\hbar-\delta)^{\mathsf T}
 \Gamma^{-1}(r_\alpha/\sqrt\hbar-\delta)]}{\pi^M\sqrt{\det\Gamma}}.
 \label{eq:b-gaussian-q}
\end{equation}
The $I/2$ contribution is the vacuum width of the {coherent state} projector; no purity assumption is required.

Introduce
\begin{equation}
 J=\begin{pmatrix}0&I\\I&0\end{pmatrix},\qquad
 T=\frac1{\sqrt2}\begin{pmatrix}I&I\\iI&-iI\end{pmatrix}.
 \label{eq:b-transforms}
\end{equation}
On the physical diagonal, $T\boldsymbol\xi=r_\alpha/\sqrt\hbar$ and $\boldsymbol\xi^{\mathsf T}J\boldsymbol\xi/2=\|\boldsymbol\alpha\|^2$. Substitution into Eq.~\eqref{eq:b-q-kernel}, followed by extension to independent variables, gives
\begin{equation}
 \mathcal K_\rho(\boldsymbol z,\boldsymbol w)
 =c_\rho\exp\left(\tfrac12\boldsymbol\xi^{\mathsf T}B\boldsymbol\xi+b^{\mathsf T}\boldsymbol\xi\right),
 \label{eq:b-gaussian-kernel}
\end{equation}
with
\begin{equation}
 \begin{aligned}
 B&=J-T^{\mathsf T}\Gamma^{-1}T,\qquad b=T^{\mathsf T}\Gamma^{-1}\delta,\\
 c_\rho&=\exp[-\delta^{\mathsf T}\Gamma^{-1}\delta/2]/\sqrt{\det\Gamma}.
 \end{aligned}
 \label{eq:b-real-abc}
\end{equation}
Ordinary transposes appear because the kernel is holomorphic in independent complex variables. The scalar $c_\rho=\langle\boldsymbol0|\rho|\boldsymbol0\rangle$ is the vacuum probability.

For computation, define the complex Husimi covariance and mean
\begin{equation}
 \begin{gathered}
 W=\frac1{\sqrt2}\begin{pmatrix}I&iI\\I&-iI\end{pmatrix},\\
 \mathsf Q=W\Gamma W^\dagger,\qquad \beta=W\delta.
 \end{gathered}
 \label{eq:b-complex-q}
\end{equation}
The matrix $\mathsf Q$ is distinct from the function $Q_\rho$. Using $T=W^\dagger J$ and $T^{\mathsf T}=W$ gives
\begin{equation}
 \begin{aligned}
 B&=(I-\mathsf Q^{-1})J,\qquad b=\mathsf Q^{-1}\beta,\\
 c_\rho&=\frac{\exp[-\tfrac12\beta^\dagger\mathsf Q^{-1}\beta]}{\sqrt{\det\mathsf Q}}.
 \end{aligned}
 \label{eq:b-complex-abc}
\end{equation}
The opposite variable-block ordering used in Ref.~\cite{Motamedi2026Stellar} is related by $B=JA_\rho^{\mathrm{ref}}J$ and $b=Jb_\rho^{\mathrm{ref}}$, with the same scalar prefactor. For a pure Gaussian state, $B=\operatorname{diag}(A,A^*)$ and $b=(u,u^*)^{\mathsf T}$. One may therefore use $F_\psi(\boldsymbol z)=c_\psi\exp(\boldsymbol z^{\mathsf T}A\boldsymbol z/2+u^{\mathsf T}\boldsymbol z)$, where $|c_\psi|^2=c_\rho$. The implementation chooses $c_\psi=\sqrt{c_\rho}$, fixing an irrelevant global phase.

\subsection{PNR conditioning, coefficient recurrence, and truncation}

Let $o$ denote the retained mode and $H$ all measured modes. For ideal PNR detection with outcome $\boldsymbol h=(h_j)_{j\in H}$,
\begin{equation}
 \widetilde\rho_o={}_H\langle\boldsymbol h|\rho|\boldsymbol h\rangle_H,
 \qquad \widetilde\rho_{nm}=\rho_{(n,\boldsymbol h),(m,\boldsymbol h)},
 \label{eq:b-conditional}
\end{equation}
where indices are reordered to place the retained mode first. Equivalently,
\begin{equation}
 \widetilde{\mathcal K}_o=
 \left.\frac{\partial_{\boldsymbol z_H}^{\boldsymbol h}\partial_{\boldsymbol w_H}^{\boldsymbol h}}
 {\prod_{j\in H}h_j!}\mathcal K_\rho\right|_{\boldsymbol z_H=\boldsymbol w_H=0},
 \label{eq:b-herald-kernel}
\end{equation}
followed by
\begin{equation}
 \widetilde\rho_{nm}=\left.\frac{\partial_{z_o}^n\partial_{w_o}^m\widetilde{\mathcal K}_o}
 {\sqrt{n!m!}}\right|_{z_o=w_o=0}.
 \label{eq:b-output-elements}
\end{equation}
Each measured mode contributes a factor $\sqrt{h_j!}$ on each side of the density matrix, explaining the denominator in Eq.~\eqref{eq:b-herald-kernel}. Such projections generally produce non-Gaussian conditional states~\cite{Su2019Conversion}.

The implementation uses coefficient recurrences rather than explicit symbolic differentiation~\cite{Miatto2020Fast}. For $\mathcal K=\sum_{\boldsymbol k}g_{\boldsymbol k}\boldsymbol\xi^{\boldsymbol k}/\sqrt{\boldsymbol k!}$, the Gaussian kernel obeys
\begin{equation}
 g_{\boldsymbol k+\boldsymbol e_i}=
 \frac{b_i g_{\boldsymbol k}+\sum_j B_{ij}\sqrt{k_j}\,g_{\boldsymbol k-\boldsymbol e_j}}
 {\sqrt{k_i+1}},\qquad g_{\boldsymbol0}=c_\rho,
 \label{eq:b-recurrence}
\end{equation}
where negative-index terms vanish. The same recurrence applies to pure-state amplitudes with the corresponding pure-state kernel parameters.

In the {density matrix} implementation, the measured-variable derivatives are first assembled as a finite polynomial:
\begin{equation}
 \widetilde{\mathcal K}_o
 =c_\rho P_{\boldsymbol h}(z_o,w_o)
 e^{\boldsymbol\xi_o^{\mathsf T}B_{oo}\boldsymbol\xi_o/2+b_o^{\mathsf T}\boldsymbol\xi_o},
 \label{eq:b-polynomial}
\end{equation}
where $\boldsymbol\xi_o=(z_o,w_o)^{\mathsf T}$ and $B_{oo},b_o$ are the corresponding output-variable blocks. The polynomial includes the factorial normalization in Eq.~\eqref{eq:b-herald-kernel} and has total degree at most $2\sum_jh_j$. Its coefficients are computed recursively and combined with those of the two-variable Gaussian factor to obtain the requested output block. This avoids constructing the full multimode Fock density tensor, although the computational cost still depends on the heralding pattern and output cutoff.

The heralding probability and normalized conditional state are
\begin{equation}
 P_{\boldsymbol h}=\operatorname{Tr}\widetilde\rho_o,\qquad
 \rho_o=\widetilde\rho_o/P_{\boldsymbol h},\quad P_{\boldsymbol h}>0.
 \label{eq:b-probability}
\end{equation}
To avoid {output truncation} bias in the probability, we compute it independently from the measured Gaussian marginal:
\begin{equation}
 P_{\boldsymbol h}=\langle\boldsymbol h|\rho_H|\boldsymbol h\rangle,
 \qquad \rho_H=\operatorname{Tr}_o\rho.
 \label{eq:b-marginal}
\end{equation}
In the $xxpp$ ordering, the marginal moments are the subvector and principal covariance submatrix selected by the measured $x,p$ indices. Reconstructing their Bargmann kernel and extracting its $(\boldsymbol h,\boldsymbol h)$ coefficient yields $P_{\boldsymbol h}$ independently of the output cutoff. Setting output variables of the full kernel to zero would instead implement a vacuum projection, not a partial trace.

For $\Pi_C=\sum_{n=0}^{C-1}|n\rangle\langle n|$, define the subnormalized truncated block and its missing probability mass by
\begin{equation}
 \rho_o^{(C)}=\frac{\Pi_C\widetilde\rho_o\Pi_C}{P_{\boldsymbol h}},
 \qquad \epsilon_C=1-\operatorname{Tr}\rho_o^{(C)}.
 \label{eq:b-tail}
\end{equation}
The block is not subsequently renormalized to unit trace. The covariance-based interface returns an unnormalized conditional block by default; when normalization is requested, it divides density matrices by $P_{\boldsymbol h}$ and pure-state amplitudes by $\sqrt{P_{\boldsymbol h}}$, preserving the truncation deficit.

For a normalized pure target $|G\rangle$, we use the {squared fidelity} convention
\begin{equation}
 F=\langle G|\rho_o|G\rangle
 =\langle G|\widetilde\rho_o|G\rangle/P_{\boldsymbol h}.
 \label{eq:b-fidelity}
\end{equation}
{Finite cutoff} evaluation uses $\Pi_C|G\rangle$ and the corresponding output block, requiring convergence checks for both target and output tails. Renormalizing the projected target changes the comparison to that normalized truncated target and must be stated explicitly.

{For the loss-aware results, heralding probabilities are obtained from the {measured mode} marginal and stabilizer expectations from the conditional characteristic function, independently of the {retained mode} cutoff. Verification compares $C=80$ and $100$, increasing the cutoff further until the {conditional state} trace deficit and absolute changes in the verified metrics are below $10^{-7}$. Neither the conditional output block nor the projected target is renormalized to remove truncation errors.}

\subsection{Numerical validation}

{The Bargmann implementation was validated in double precision against analytical and independent numerical references. Benchmarks using vacuum, coherent, squeezed-vacuum, and thermal states confirmed agreement with analytical Fock coefficients under the adopted kernel conventions. The pure-state limit was verified through the relation $\rho_{nm}=\psi_n\psi_m^*$. For a single-mode state undergoing displacement, squeezing, and optical loss, the full complex {density matrix} block was compared with an independent {finite Fock space} calculation using matrix exponentials and loss Kraus operators. The maximum absolute {matrix element} deviation was approximately $7.3\times10^{-15}$. For a displaced three-mode mixed state, joint PNR probabilities agreed with DeepQuantum to a maximum absolute deviation of approximately $9.4\times10^{-17}$ within the tested output block.}

{Automatic derivatives with respect to displacement, squeezing, and loss were checked against central finite differences, including at zero displacement. Additional tests checked the quadrature conventions and the independence of the marginal heralding probability from the {retained mode} cutoff. These comparisons establish numerical consistency for the tested cases. The accuracy of the reported circuit results is assessed separately through the {conditional state} trace deficit and convergence of the evaluated observables, as described above.}

\section{Optical loss as coupling to a vacuum environment}
\label{app:optical-loss}

\begin{figure}[t]
 \centering
 \includegraphics[width=\linewidth]{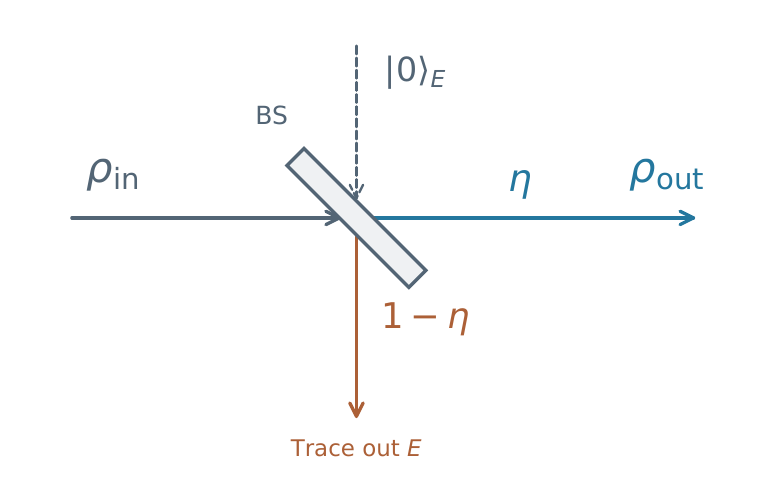}
 \caption{Beam-splitter representation of optical pure loss. The signal input $\rho_{\mathrm{in}}$ mixes with a vacuum environment at a BS of intensity transmission $\eta$. Tracing out the environment output yields the attenuated signal state $\rho_{\mathrm{out}}=\mathcal E_\eta(\rho_{\mathrm{in}})$. The dashed input denotes the vacuum environment; the downward output is traced out. The labels $\eta$ and $1-\eta$ denote the signal transmission and loss fractions, respectively.}
 \label{fig:loss-model}
\end{figure}

Each optical loss channel is modeled by mixing the signal with an independent vacuum environment on a beam splitter (BS) and tracing out the environment output~\cite{Weedbrook2012Gaussian}, as illustrated in Fig.~\ref{fig:loss-model}. For intensity transmission $\eta$, the signal operator transforms as
\begin{equation}
 \hat a_{\mathrm{out}}=\sqrt{\eta}\,\hat a_{\mathrm{in}}
 +\sqrt{1-\eta}\,\hat e_{\mathrm{in}},
 \label{eq:c-bs-loss}
\end{equation}
where $\hat e_{\mathrm{in}}$ is the vacuum environment mode. The resulting pure loss channel is
\begin{equation}
 \mathcal E_\eta(\rho)=\operatorname{Tr}_E\!\left[
 U_{\mathrm{BS}}(\eta)\bigl(\rho\otimes|0\rangle_E\langle0|\bigr)
 U_{\mathrm{BS}}^\dagger(\eta)\right].
 \label{eq:c-loss-channel}
\end{equation}
The environment output is unobserved; no measurement or postselection is applied to it. An attenuation of $\ell$ decibels corresponds to $\eta=10^{-\ell/10}$, so $1\,\mathrm{dB}$ gives $\eta\approx0.7943$. The corresponding amplitude transmission is $\sqrt{\eta}$.

In the quadrature convention of Appendix~\ref{app:bargmann}, this channel maps the single-mode moments to $\boldsymbol d'=\sqrt{\eta}\,\boldsymbol d$ and $\Sigma'=\eta\Sigma+(1-\eta)\hbar I_2/2$. The second term accounts for vacuum noise entering through the unused BS port. Equation~\eqref{eq:b-loss} implements the corresponding multimode map directly without environmental Fock truncation, with independent vacuum environments applied at the respective channel locations specified in Appendix~\ref{app:spatial-loss-circuits}.

\FloatBarrier

\bibliography{main}

\end{document}